\documentclass{sig-alternate-05-2015}

\usepackage{multirow}
\usepackage[show]{chato-notes}

\usepackage{framed}
\usepackage{balance}
\usepackage{multirow}
\usepackage{graphics}
\usepackage{url}
\usepackage{wrapfig}
\usepackage{algorithmic} 
\usepackage[]{algorithm2e} 
\usepackage{float} 
\usepackage{amsmath}
\usepackage{amssymb}
\usepackage{xspace}
\usepackage{amsfonts}

\numberwithin{equation}{section}

\def\pprw{8.5in}
\def\pprh{11in}

\usepackage{subfig, float, epsfig, epstopdf, balance}
\usepackage{helvet}
\usepackage{courier}
\usepackage{multirow}
\usepackage{mathtools}

\usepackage{color, soul}
\usepackage[table]{xcolor}
\usepackage{booktabs}
\definecolor{lightgray}{gray}{0.85}

\usepackage{amsmath}
\def\P{\mathbb{P}}

\newcommand{\bv}{\mathbf{v}}

\usepackage{caption} 
\title{Scalable Semantic Matching of Queries to Ads \\ in Sponsored Search Advertising}

\numberofauthors{2} 
\author{
\alignauthor Mihajlo Grbovic, Nemanja Djuric, \\ Vladan Radosavljevic, Fabrizio Silvestri, Ricardo Baeza-Yates \\
\affaddr{Yahoo Labs, Sunnyvale, CA, USA} \\
\email{\{mihajlo, nemanja, vladan, silvestr, rby\}@yahoo-inc.com}
\alignauthor Andrew Feng, Erik Ordentlich, \\ Lee Yang, Gavin Owens\\
\affaddr{Yahoo Hadoop Group, Sunnyvale, CA, USA} \\
\email{\{afeng, eord, leewyang, gowens\}@yahoo-inc.com}
}

\begin{document}

\CopyrightYear{2016} 
\setcopyright{acmcopyright}
\conferenceinfo{SIGIR '16,}{July 17-21, 2016, Pisa, Italy}
\isbn{978-1-4503-4069-4/16/07}\acmPrice{\$15.00}
\doi{http://dx.doi.org/10.1145/2911451.2911538}

\maketitle{}
\RestyleAlgo{boxed}

\begin{abstract}
Sponsored search represents a major source of revenue for web search engines. This popular advertising model brings a unique possibility for advertisers to target users' immediate intent communicated through a search query, usually by displaying their ads alongside organic search results for queries deemed relevant to their products or services. However, due to a large number of unique queries it is challenging for advertisers to identify all such relevant queries. For this reason search engines often provide a service of advanced matching, which automatically finds additional relevant queries for advertisers to bid on. We present a novel advanced matching approach based on the idea of semantic embeddings of queries and ads. The embeddings were learned using a large data set of user search sessions, consisting of search queries, clicked ads and search links, while utilizing contextual information such as dwell time and skipped ads. To address the large-scale nature of our problem, both in terms of data and vocabulary size, we propose a novel distributed  algorithm for training of the embeddings. Finally, we present an approach for overcoming a cold-start problem associated with new ads and queries. We report results of editorial evaluation and online tests on actual search traffic. The results show that our approach significantly outperforms baselines in terms of relevance, coverage, and incremental revenue. Lastly, we open-source learned query embeddings to be used by researchers in computational advertising and related fields.

\end{abstract}

\terms{Information Retrieval, Algorithms, Experimentation.}
\keywords{Sponsored search; ad retrieval; word embeddings.} 

\section{Introduction}
\label{sec:introduction}
The remarkable growth of the Internet in the last decades has brought huge benefits to its users, allowing easy, one-click access to all kinds of information and services. However, along with the increase in size and variety of the web there inevitably came the increase in its complexity, which left many users lost in the immense digital jungle of web content. To mitigate the situation, very early in the web's existence there appeared search engines such as Altavista, AOL, or Lycos, that allowed users to easily browse and discover webpages of interest. The service has evolved and improved significantly since these beginnings, and today search engines represent one of the most popular web services. 

Advertisers are interested in making use of the vast business potential that the search engines offer. Via issued search query users communicate a very clear intent that allows for effective ad targeting. This idea is embodied in the sponsored search model \cite{jansen2008sponsored}, where advertisers {\it sponsor} the top search results in order to redirect user's attention from original (or {\it organic}) search results to ads that are highly relevant to the entered query. Sponsored search drives significant portions of traffic to websites, and accounted for $46\%$ of overall online advertising spend of astonishing $\$121$ billion in 2014 alone\footnote{zenithoptimedia.com, accessed January 2016}. Moreover, with the advent of handheld devices the sponsored search also made a successful transition to mobile platforms. Here, it equally shares the advertising market with display ad formats, and is projected to reach $\$12.85$ billion spend in 2015\footnote{emarketer.com, accessed January 2016}.

The booking process for ad campaigns in sponsored search is typically self-served. This means that the advertisers create their own ads by providing ad creative to be shown to the users (comprising concise title, description, and display URL), along with ancillary ad parameters that are visible only to the web search publisher. These include the list of bid terms (i.e., queries for which advertisers wish to show their ad) and their bid values (i.e., monetary amounts they are willing to pay if the ad is shown and clicked). In the model currently used by most major search engines, in a case that multiple advertisers are competing for the same query the selected ads enter a generalized second price (GSP) auction \cite{edelman2005internet}, where the winner pays the runner-up's bid value when a user clicks on the shown ad. 

\begin{table}[t]
\footnotesize
\caption{Examples of variant and broad match queries}
\label{tbl:example_ad}
\begin{center}
\begin{tabular}{ l l}
\hline
\rowcolor{lightgray}
{\bf Ad title} & Host a Fun Murder Mystery Party \\
\hline
& Huge selection of fun murder mystery \\ 
{\bf Ad} & games for all ages, groups. \#1 site for \\ 
{\bf description} & instant downloads and boxed sets of \\ & exciting murder mystery party games! \\
\hline
\rowcolor{lightgray}
{\bf Bid term} & murder mystery party \\
\hline
{\bf Variant} & murder mystery parties \\
{\bf matches} & murder mistery party \\
\hline
\rowcolor{lightgray}
& murder mystery dinner \\
\rowcolor{lightgray}
& murder mystery parties at home \\
\rowcolor{lightgray}
{\bf Broad} & murder mystery dinner party \\
\rowcolor{lightgray}
{\bf matches} & how to host a mystery party \\
\rowcolor{lightgray}
& friends game night \\
\rowcolor{lightgray}
& detective games \\
\rowcolor{lightgray}
& murder mystery games for parties\\
\rowcolor{lightgray}

\hline
\end{tabular}
\end{center}
\end{table}

The principal strategy for ad selection is based on explicit matching of user search queries to all the bid terms in the system, referred to as {\it exact match}. Provided that the advertiser-defined bid terms are indeed relevant to the ad, exact match ensures that advertisers interact only with the relevant audience. However, as it is nearly impossible for advertisers to explicitly list all commercially relevant bid terms, it may also limit user reach and lead to lower traffic volume for the advertiser \cite{goel2010anatomy}. This can clearly result in lost opportunities and revenue for both the search engine and the advertisers. To address this problem a service of {\it advanced matching} is usually offered to the advertisers.

More specifically, with advertiser's permission search engines attempt to match ads to related queries not explicitly provided by the advertisers, by implementing {\it variant} and {\it broad} match techniques (illustrated in Table \ref{tbl:example_ad}). Variant match looks for slight variations of the bid term (e.g., typos, inflections), typically by leveraging Levenshtein distance \cite{levenshtcin1966binary}. Broad match is a more involved process of matching queries to advertiser-provided bid terms that do not necessarily have any words in common, yet have same or similar meaning (e.g., query ``running machine" and bid term ``elliptical trainer"). Existing techniques include expanding user query to find similar queries that can be matched exactly to existing bid terms, commonly referred to as query rewriting \cite{boldi08queryflow,grbovic2015context}, or computing similarity between query and text in ad title and description. Query rewriting approaches are limited to matching only against defined bid terms and may not be suitable in cases when advertisers do not provide relevant bid terms to begin with. On the other hand, matching queries to ads based solely on textual information is problematic in case when no common words can be found.

In this paper we describe a novel broad match technique that was recently deployed in sponsored search system of Yahoo Search, which effectively and efficiently addresses the shortcomings of the existing broad match techniques. Motivated by recent advances in distributed neural language models \cite{bengio2006neural,collobert2011natural,mikolov2013distributed,turian2010word}, we aim at learning low-dimensional continuous representations of queries and ads in a common vector space. To do so we rely on the search sessions, defined as sequences of queries, ad clicks and search link clicks performed by a particular user, and leverage surrounding context of queries and ads in a session to learn vectors that best describe them. As a result ads and queries which occurred in similar contexts will have similar representations, and finding relevant ads for a specific query amounts to a simple $K$ nearest neighbor search in the learned embedding space, without ever considering the text in queries and ads.

Compared to our previous work \cite{grbovic2015context}, novel contributions of this paper are summarized below:
\begin{itemize} 
\itemsep0em
\item {\bf Implicit negatives}. We incorporated implicit negative signal (in terms of skipped, non-clicked ads) into the training procedure, leading to higher quality broad matches compared to training with random negatives;
\item {\bf Dwell time weights}. We integrated dwell time (i.e., time spent on the advertiser's page following an ad click) into the training procedure, allowing us to weight short and long clicks differently, according to a strength of the interest signal;
\item {\bf Large-scale training}. To address the scalability problems due to training data size and the need for large vocabulary (amounting to hundreds of millions of unique queries and ads), we proposed a novel distributed algorithm for training embeddings using parameter server framework that results in 10x training time speedup and over 5x bigger vocabulary compared to single machine training described in \cite{mikolov2013distributed};
\item {\bf Cold-start ad embeddings}. We solved an important problem of computing embeddings for ads that were not observed in training, such that newly booked ads can be instantly matched to user queries;
\item {\bf Cold-start query embeddings}. We proposed a procedure for calculating high-quality embeddings for new and tail queries using embeddings of head queries;
\item {\bf Experiments on real search traffic}. We provide insights on how different state-of-the-art algorithms perform when applied to actual search traffic;
\item {\bf Open-source vectors}. As part of this work we open-sourced a portion of query embeddings for research purposes.
\end{itemize} 

We trained a search embedding model using more than $9$ billion search sessions, resulting in ad and query representations of very high quality. Extensive evaluation on real-world search traffic showed that the proposed approach significantly outperformed the existing state-of-the-art broad match methods on a number of critical performance measures, including relevance, reach, and click-through rate (CTR). 

\section{Related work}
In the following we introduce related work on advanced query-ad matching in sponsored search, and recent advances in neural language models and distributed embeddings.

\subsection{Neural language models}
It has been shown that a number of problems in Natural Language Processing (NLP) domain can be effectively solved by defining a probability distribution over word sequences, shown to perform well in chunking, parsing, or sentiment analysis, to name a few. For this reason researchers have proposed a number of methods called language models to mathematically represent generation of language, aimed to capture statistical characteristics of written language \cite{baeza1999modern,lavrenko2001relevance}. Classic methods for language modeling represented words as high-dimensional, sparse vectors using one-hot representation. Using this approach word vectors have dimensionality equal to the size of the entire vocabulary, with zero values everywhere except for the element corresponding to the given word. However, the approach often gives suboptimal results due to sparsity issues and curse of dimensionality.

In order to mitigate the problems with word representations in high-dimensional spaces, neural language models have been introduced. Unlike classical approaches, these methods learn distributed, low-dimensional representations of words through the use of neural networks \cite{bengio2006neural,collobert2011natural,turian2010word}. The networks are trained by directly taking into account the word order and their co-occurrence, based on the assumption that words frequently appearing together in the sentences also share more statistical dependence.

Despite their effectiveness, training a large number of parameters of neural network-based approaches has been a serious obstacle to wider use of the neural models. In particular, a single parameter update requires iteration over the entire vocabulary, which can grow prohibitively large for many practical applications. However, with the recent advances in the NLP domain, and in particular with the development of highly scalable continuous bag-of-words (CBOW) and skip-gram (SG) language models for word representation learning \cite{mikolov2013distributed}, the models have been shown to obtain state-of-the-art performance on many traditional language tasks after training on large-scale textual data sets. 

\subsection{Broad match in sponsored search}
Broad match is a well established, often used approach in sponsored search advertising, responsible for billions of dollars of search engine revenue. In addition to exact and variant match, which typically have the highest click-through rates, many advertisers use broad match to increase the reach to relevant queries that were not explicitly provided, entrusting search engines with this task.

Most of the existing broad match techniques rely on query rewriting to rewrite an unmatched user query to several similar queries, in expectation that one of the rewrites will match an advertiser-provided bid term, resulting in retrieval of ads that bidded on that term. A basic query rewriting approach involves representing queries as bag-of-words using tf-idf weighting, and calculating similarity between the query and all other queries in order to find good rewrite candidates. However, the bag-of-words representation of queries is extremely sparse, which makes it impossible to find related queries that do not share common words. Moreover, it can lead to false positives when queries share common words that have different meanings in different contexts, e.g. ``house music" and ``house prices". To mitigate these issues some researchers proposed to enrich query representation using text from web search results \cite{broder2008search}. Others proposed to extend query rewriting beyond string matching by applying graph-based methods, such as Query Flow Graph (QFG) \cite{boldi08queryflow}, where similar queries are found by a random walk over a bipartite graph of queries and link clicks \cite{fuxman2008using}, or neural networks \cite{grbovic2015context}, where query feature representations are learned from web search data. However, none of these methods can overcome an inherent limitation of the query rewriting approach. More specifically, once the query is rewritten it can only be matched to ads that have the resulting rewrites as bid terms, potentially overlooking other relevant ads. 

Alternative methods aim to address this issue by directly retrieving ads by calculating similarity between bag-of-words representation of queries and bag-of-words representation of ads obtained from ad title and description. Even though it reduces sparsity, this approach does not completely solve the vocabulary mismatch problem \cite{ribeiro2005impedance} and may even introduce additional issues, especially in the case of {\it templated} ad creatives where majority of phrases in titles and descriptions have little to do with the advertised product (e.g., ``free shipping", ``best prices", ``buy now").

Finally, to rank the ads by relevance (e.g., in the case too many ads are retrieved via query rewriting), it is common to apply the learning-to-rank approach \cite{hang2011short} that learns weights for query and ad features based on editorially-labeled training set. These methods typically require enormous amounts of accurate editorial grades for (query, ad) matches that are often very expensive to obtain.

In this paper we go beyond query rewriting and supervised learning-to-rank, and propose to learn query and ad embeddings from search data in an unsupervised manner, followed by directly matching and ranking ads for queries based on the distances in the learned embedding space. 

\section{Methodology}
To address the shortcomings of existing broad match algorithms in sponsored search, we propose to take a new approach to this task, motivated by the recent success of distributed language models in NLP applications \cite{mikolov2013distributed,turian2010word}. In the context of NLP, distributed models are able to learn word representations in a low-dimensional continuous vector space using a surrounding context of the word in a sentence, where in the resulting embedding space semantically similar words are close to each other \cite{mikolov2013distributed}. Our objective is to take advantage of this property for the task of query-ad matching in sponsored search, and to learn query and ad representations in a low-dimensional space where queries would be close to related ads. This would allow direct matching of queries to ads, instead of taking the longer route of query rewriting. Clearly, such approach reduces the complex broad match task to a trivial $K$-nearest-neighbor ($K$-nn) search between queries and ads in the joint embedding space.

Finding distributed query and ad representation, as opposed to finding word representations, brings very unique challenges. For example, while the basic unit for learning word representations is a sentence $s = (w_{1}, \ldots, w_{M})$ consisting on $M$ words $w_m, m = 1,\ldots,M$, in our proposed approach the basic unit for learning query and ad representations are user actions within a search session $s = (a_{1}, \ldots, a_{M})$, where in the simplest case an action can be a search query or an ad click. Moreover, search sessions have a number of additional contexts that can be used to improve quality of the final model. For instance, search sessions typically contain link clicks. Even though we are only interested in query and ad embeddings, link clicks can provide additional context. In essence, the query and ad embeddings will be affected by the co-click behavior between queries and search link clicks, resulting in an improved model. In addition, ad dwell time can be leveraged to distinguish between good ad clicks and unsatisfactory or accidental clicks \cite{lalmas2015promoting}, and can provide useful context for training better quality embeddings. Finally, ads that are skipped in favor of a click on a lower positioned ad can be used as implicit negative signal and serve as negative context during training to improve the resulting model.

To formalize the training procedure, let us assume we are given a set $\mathcal{S}$ of $S$ search sessions obtained from $N$ users, where each session $s = (a_{1}, \ldots, a_{M}) \in \mathcal{S}$ is defined as an uninterrupted sequence of $M$ user actions comprising queries, ad clicks, and link clicks. A new session is initiated whenever there is a time gap of more than 30 minutes between two consecutive user actions \cite{gayoavello09survey}. Then, the learning objective is to find $D$-dimensional real-valued representation ${\bf v}_{a_m} \in \mathbb{R}^D$ of each unique user action $a_m$.

The proposed search embedding model (called {\it search2vec}) learns user search action representations using the skip-gram model \cite{mikolov2013distributed} by maximizing the objective function $\mathcal{L}$ over the entire set $\mathcal{S}$ of search sessions, defined as follows,
\begin{equation} \label{cx2v_obj}
\mathcal{L} = \sum_{s \in \mathcal{S}} \sum_{a_m \in s} \sum_{-b\le i\le b, i\ne 0} \log \P(a_{m+i}|a_m).
\end{equation} 
Probability $\P(a_{m+i}|a_m)$ of observing a neighboring user action $a_{m+i}$ given the current action $a_m$ is defined using the soft-max function,
\begin{equation}\label{cx2v_prq}
\P(a_{m+i}|a_m) = \frac{\exp(\mathbf{v}_{a_m}^\top \mathbf{v}_{a_{m+i}}^\prime)}{\sum_{a=1}^{|\mathcal{V}|} \exp(\mathbf{v}_{a_m}^\top \mathbf{v}_{a}^\prime)},
\end{equation}
where $\mathbf{v}_{a}$ and $\mathbf{v}_{a}^\prime$ are the input and output vector representations of user action $a$, hyperparameter $b$ is defined as a length of the relevant context for action sequences, and $\mathcal{V}$ is a vocabulary defined as a set of unique actions in the data set comprising queries, ads, and links.
From \eqref{cx2v_obj} and \eqref{cx2v_prq} we see that search2vec models temporal context of action sequences, where actions with similar contexts (i.e., with similar neighboring actions) will have similar representations.

Time required to compute gradient $\nabla \mathcal{L}$ of the objective function in \eqref{cx2v_obj} is proportional to the vocabulary size, which may be computationally infeasible in practical tasks as $|\mathcal{V}|$ could easily reach hundreds of millions. As an alternative we used negative sampling approach proposed in \cite{mikolov2013distributed}, which significantly reduces computational complexity. Negative sampling can be formulated as follows. We generate a set $\mathcal{D}_p$ of pairs $(a,c)$ of user actions $a$ and their contexts $c$ (i.e., actions within a window of length $b$ that either precede or follow action $a$), and a set $\mathcal{D}_n$ of negative pairs $(a, c)$ of user actions and $n$ randomly sampled actions from the entire vocabulary. The optimization objective then becomes,
\begin{equation}\label{cx2v_neg}
\underset{\theta}{\operatorname{argmax}} \sum_{(a,c) \in \mathcal{D}_p} \log \frac{1}{1+ e^{- \mathbf{v}'_{c} \mathbf{v}_{a}}} + \sum_{(a,c) \in \mathcal{D}_n} \log \frac{1}{1+ e^{\mathbf{v}'_{c} \mathbf{v}_{a}}},
\end{equation}
where parameters $\theta$ to be learned are $\mathbf{v}_{a}$ and $\mathbf{v}'_{c}$, $a, c \in \mathcal{V}$. The optimization is done via stochastic gradient ascent. 

\begin{figure}[t]
\centering
\includegraphics[width=0.41\textwidth]{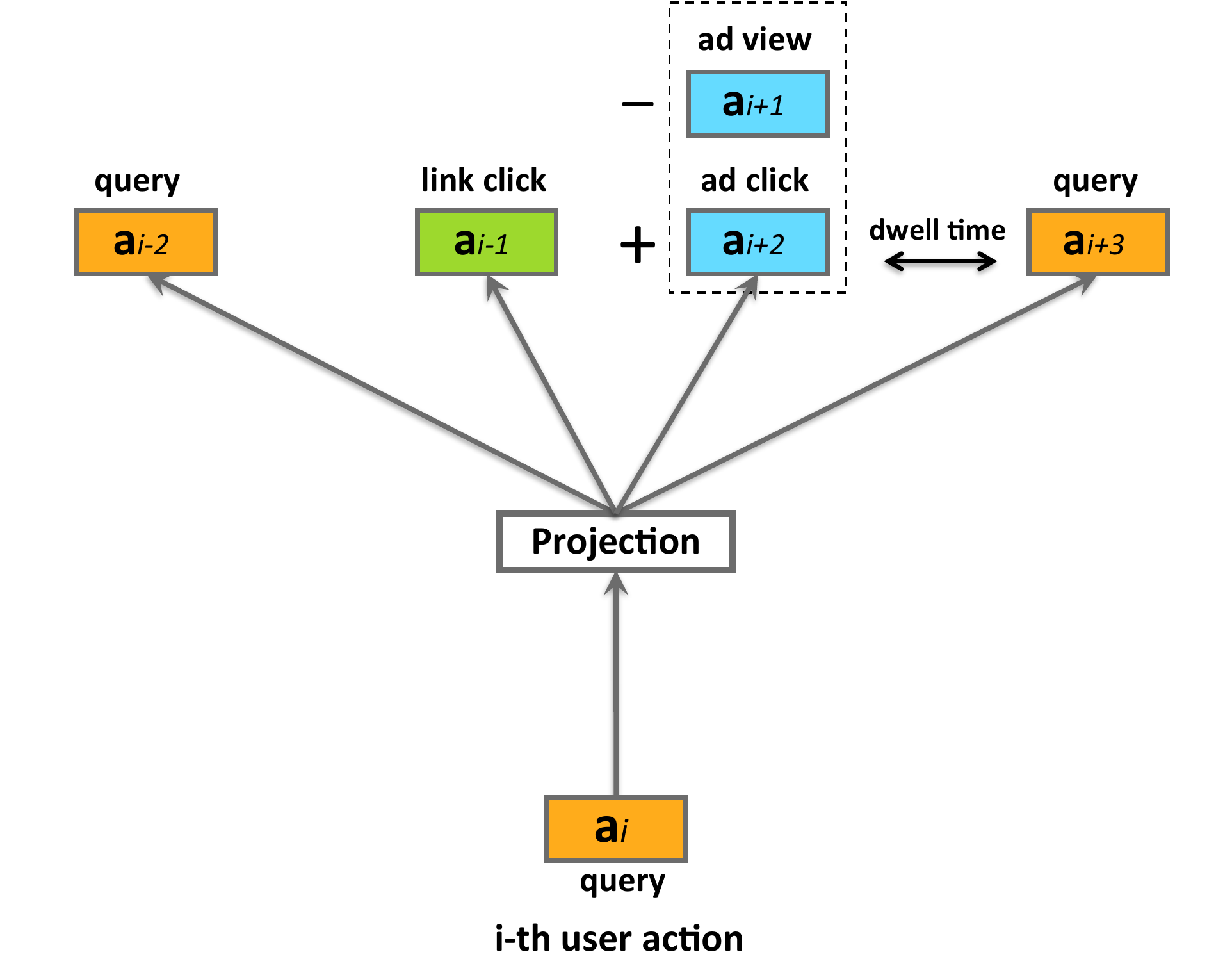}
\caption{search2vec: dwell-time implicit negative skip-gram}
\label{fig:antares_skip}
\end{figure}

Visual representation of the search2vec model is presented in Figure~\ref{fig:antares_skip}. As discussed previously, and as depicted in Figure~\ref{fig:antares_skip}, we propose to utilize additional context specific to sponsored search domain, namely implicit negatives and dwell-time. In the experimental section we explore this approach and empirically evaluate their benefits in detail.

{\bf Leveraging ad dwell time.} Let us assume that a user performed a search query $q_1$, followed by an ad click $ad_{1}$. We define {\it ad dwell time} as the time that the user spent on the ad landing page. When the user stayed longer on the ad landing page, it is usually interpreted as a positive engagement signal, as opposed to the user immediately bouncing back from the advertiser's website. There also exists a notion of accidental ad clicks with dwell time of only a few seconds, which is especially apparent on mobile phone devices. To incorporate this signal into our learning framework we propose to weight query-ad pairs with higher dwell times more, while penalizing shorter dwell times.

More specifically, when optimizing the objective function in \eqref{cx2v_neg}, we introduce an additional weight $\eta_t = \log (1+t)$ for each element of the first sum, where $t$ denotes the dwell time in minutes. For dwell times above $10$ minutes we set $\eta_t=1$. Moreover, we use this approach to compute weight only when updating the ad vector with the immediately preceding query and when updating the query vector with the immediately succeeding ad, otherwise we set $\eta_t=1$.

{\bf Leveraging implicit negative signals.} Unlike in the NLP domain, in the domain of sponsored search there does exist an implicit negative signal in a form of ads that the user decided to skip in favor of a click on a lower positioned ad, which can be utilized during training. In particular, given an issued user query $q_1$, let us assume that after the GSP auction a total of $k$ ads, $ad_{1}, ad_{2}, \ldots, ad_{k}$, were shown at different positions on the page, with $ad_{1}$ being shown at the top position in the page followed by the remaining $k-1$ ads below. Note that the top position has the highest value to the advertisers, as it has been shown to have the highest probability of being clicked, especially on mobile devices due to limited screen sizes. We adopt an assumption that if the user scrolls down the page and clicks on a lower positioned ad, all the ads ranked higher were of no interest to him and can be treated as negative signal for query $q_1$. We consider implicit negatives only on top $k=3$ positioned ads, and only in cases where user had a single ad click in the session with dwell time of over $10$ seconds.

We formulate the use of implicit negatives in the following way. In addition to sets $\mathcal{D}_p$ and $\mathcal{D}_n$, we generate a set $\mathcal{D}_{in}$ of pairs $(q,ad)$ of user queries $q$ and the ads $ad$ at higher positions which were shown to the user, and skipped over in favor of a click on ad at one of the lower positions. The new optimization objective then becomes as follows,
\begin{equation} \label{cn2v_eneg}
\begin{aligned}
\underset{\theta}{\operatorname{argmax}} & \sum_{(a,c) \in \mathcal{D}_p} \log \frac{1}{1+ e^{- \mathbf{v}'_{c} \mathbf{v}_{a}}} + \sum_{(a,c) \in \mathcal{D}_n} \log \frac{1}{1+ e^{\mathbf{v}'_{c} \mathbf{v}_{a}}} \\
& + \sum_{(q,ad) \in \mathcal{D}_{in}} \log \frac{1}{1+ e^{\mathbf{v}'_{ad} \mathbf{v}_{q}}},
\end{aligned}
\end{equation}
where parameters $\theta$ to be learned remain the same as before.

{\bf Inference.} Given learned query and ad embeddings, ad retrieval for a given query $q$ is performed by calculating cosine similarity between its vector ${\bf v}_q$ and all ad vectors ${\bf v}_{ad}$ in the vocabulary. The $K$ ads with the highest similarity are retrieved as broad matches. For large vocabularies, efficient $K$-nn search can be done via locality-sensitive hashing \cite{gionis1999similarity}.

\subsection{Scalable training algorithm}
\label{sec:scalable2vec}
To realize the full potential of the proposed approach, we found it necessary to train embeddings for several hundred million of actions (i.e., {\it words} in the ``vocabulary''), which include queries, ads, and links comprising the search session data. Existing implementations for training word embeddings were found to fall short for our vocabulary size target, as they require that all of the input and output vectors fit in the memory of a single machine. For example, a vocabulary of 200 million words with 300-dimensional vectors requires 480GB of RAM memory, which is beyond the capacity of typical machines. To address this issue, we developed a novel distributed word embedding training algorithm based on a variation of the parameter server paradigm.

A parameter server (PS) ~\cite{dean2012parameterserver} is a high-performance, distributed, in-memory key-value store specialized for machine learning applications, typically used to store latest values of model parameters throughout the course of training. A PS-based training system includes a number of PS shards that store model parameters and a number of clients that read different portions of the training data from a distributed file system (e.g., Hadoop Distributed File System). 

In the conventional PS system, the word embedding training can be implemented as follows. Each PS shard stores input and output vectors for a portion of the words from the vocabulary. Each client reads a subset of training data in minibatches (e.g., 200 lines at a time), and after each minibatch performs the vector updates of words found in the minibatch and randomly sampled negative words. The updates are achieved by transmitting the IDs of the words to all the PS shards, which will, as a result, send back the corresponding vectors to the client. The client calculates the gradient descent updates with respect to the minibatch restriction of the objective \eqref{cx2v_neg} and transmits them to the PS shards where the actual updates happen.

\begin{figure}
\centering
\includegraphics[width=2.83in]{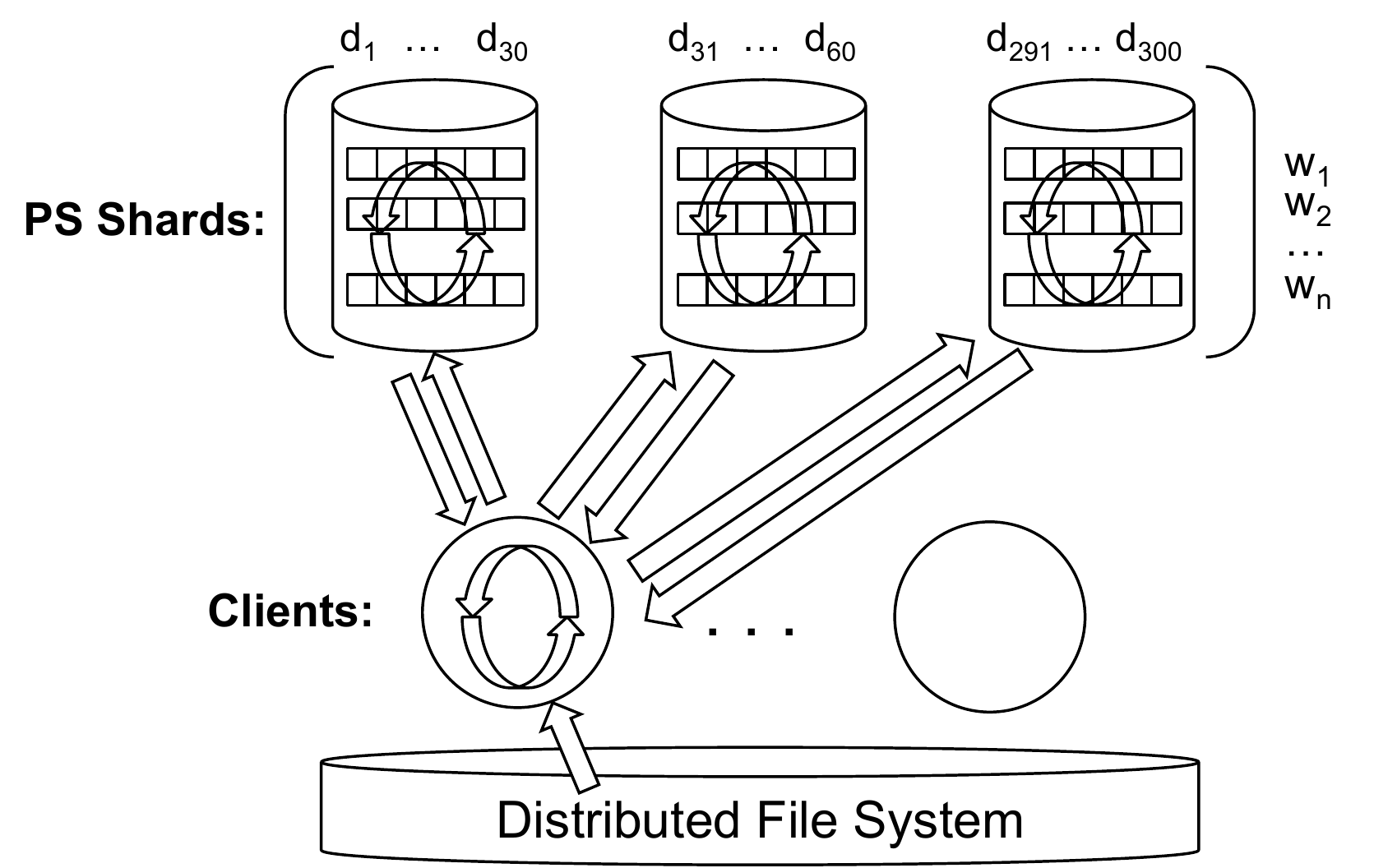}
\caption{Scalable word2vec on Hadoop}
\label{fig:w2v}
\end{figure}

However, due to transmission costs of the actual word vectors over the network, the conventional PS-based implementation of word embedding training requires too much network bandwidth to achieve acceptable training throughput. For this reason we propose a different distributed system architecture that requires significantly less network bandwidth. The proposed algorithm, illustrated in Figure~\ref{fig:w2v}, features the following contributions with respect to the conventional PS-based training approach:
\begin{itemize}
\itemsep0em
\item Column-wise partitioning of vectors among PS shards;
\item No transmission of word vectors across the network;
\item PS shard-side negative sampling and computation of partial vector dot products.
\end{itemize}

In our algorithm, each PS shard stores distinct portion of vector dimensions for {\it all} words from the vocabulary ${\cal V} $, as opposed to previously described word-wise partitioning where each PS shard stores entire vectors for a subset of words. For example, assuming a vector dimension $d = 300 $ and 10 PS shards, shard $s \in \{0,\ldots,9\}$ would store 30 dimensions of input vectors $ \bv(w) $ and output vectors $ \bv'(w) $ with indices $ i $ in the range $ 30 s \leq i < 30s + 30 $ for all vocabulary words.

Since each PS shard now stores the entire vocabulary (i.e., different vector dimensions of {\it all} words), we can implement the random sampling step and the dot product calculation step to the PS shards themselves. In this way we avoid transmission costs of the actual vectors. Each client still processes data in minibatches, and sends to the PS shards the indices of (word, context) pairs that need to be updated. In addition, the client broadcasts common random number generator seeds, such that all shards can sample the same negative context words. Each PS shard then calculates the partial dot products with respect to their local dimensions for each of the positive and negative (word, context) pairs and sends the result to a client that made the request. Finally, the client aggregates local dot products it received from all PS shards and calculates the global coefficient needed for gradient updates. The single client's minibatch step ends by sending the global coefficients to all PS shards, where input and output vectors are updated correspondingly, without any locking or synchronization mechanism.

In a typical at-scale run of the algorithm, the above process is carried out by multiple client threads running on each of a few hundred cluster nodes, all interacting with the PS shards in parallel. The dataset is iterated over multiple times, and after each iteration the learning rate $ \alpha $ is reduced. The distributed implementation handles implicit negatives in the same manner as positive contexts only with negative multiplier in the gradient update, and can handle dwell-time weights seamlessly. In comparison to~\cite{mikolov2013distributed}, our system was found to be up to 10x faster depending on the load on the shared cluster, training set, and vocabulary size.

\subsection{Cold start ad embeddings}
\label{sec:c2vec}

Contrary to everyday language where new words come into existence occasionally, in ad serving business new ads are introduced to the system every day. Matching service is required to find good query candidates for these new ads even before they are shown to users. In this section we propose a method that addresses this problem by producing cold start ad embeddings of good quality, which are then used to find relevant queries for that ad.

Given a newly booked ad, in absence of ad clicks that could be used to learn a context vector, we must rely on its textual content, including the bid term, ad title, and description. In particular, in the process of learning query context vectors $V_q$ we learned embeddings for a large number of words and phrases, many of which appear in ad titles and descriptions. In addition, the bid term context vector is likely to be available (in $80\%$ of cases by our evaluation).

The bid term context vector itself could be a good choice for the cold start ad embedding. However, even though the bid term should summarize the message being conveyed in the ad, oftentimes advertisers bid on general terms, such as ``shoes", ``fitness", or ``sports", to maximize their traffic potential. Evidently, one could find additional useful phrases in the ad title and description, whose vectors, in combination with the bid term vector, could produce an embedding that is closer to more specific query matches for that ad.

To compute a content-based ad vector we propose to extract relevant words and phrases from ad content and compute the linear combination of their context vectors,
\begin{equation} \label{content_vec}
\mathbf{v}_{ad} = \sum_{t=1:T_a} \mathbf{v}_{p_{t}},
\end{equation}
where $T_a$ is the number of extracted words and phrases with vectors $\mathbf{v}_{p_{t}}$ in $V_q$. The resulting ad vector is of the same dimensionality and in the same feature space as the query context vectors. Consequently, the same retrieval procedure as the one used for context ad vectors can be used to find relevant broad match queries for the cold-start ads.

\begin{algorithm}[t]
\footnotesize
\SetKwInOut{Input}{Input}
\KwData{ad title, ad description, ad URL, bid term $b$, threshold $\tau_c$, set of context query vectors ${\it V}_q$ }
{\bf Initialization}: $\tau_c=0.45$, get bid term vector ${\bf v}_b$ from ${\it V}_q$, set ${\bf v}_{ad} = {\bf v}_b$, create ad document ${\bf d}_{ad}$ from all $n$-grams ($n=1,..,10$) in ad title, description, URL\;
\ForAll{phrases $p$ in ad document ${\bf d}_{ad}$}{
\If{${\bf v}_p$ exists in ${\it V}_q$}{
\If{sim(${\bf v}_b$,${\bf v}_p$)$>\tau$}{
${\bf v}_{ad} = {\bf v}_{ad} + {\bf v}_p$\;
}
}{
}
}
\KwResult{ad content vector ${\bf v}_{ad}$ }
\caption{content search2vec}
\end{algorithm}

The most challenging part of this procedure is extracting relevant words and phrases from the textual ad information. For example, let us consider an ad with title ``Visit Metropolitan Opera House in New York" and bid term ``performing arts". While phrases ``opera house" and ``metropolitan opera house" are relevant in the context of this ad, other phrases such as ``visit", ``house", or ``house in new york" do not carry any useful information or have completely different meaning from what is communicated in the ad. For example, phrase ``visit" is too general and ``house in new york" refers to real-estate in New York with no connections to performing arts. In addition, ads often contain generic phrases such as ``free shipping", ``buy now", ``30\% off", and ``best deals on", which if used in the final sum \eqref{content_vec} may cause the resulting vector to diverge from relevant queries in the embedding space, leading to poor broad matches.

To decide which $n$-grams should be formed from words in short text, a typical approach is to use a CRF segmentation model \cite{lafferty2001conditional}. However, CRF models have their limitations: 1) they require a large training set with human-annotated examples of correct text segmentations, which needs to be constantly updated to account for new entities; 2) they cannot automatically detect phrases that should not be considered and would require creating a blacklist of terms that should be excluded from \eqref{content_vec}. Similar shortcomings are present in the paragraph2vec model \cite{le2014distributed} that suggests simply using all words in a document to learn its embedding. 

To this end we propose an {\it anchor phrases} model, summarized in Algorithm 1, that uses search context vectors to automatically filter out stopwords and non-relevant phrases and keep only the ones closely related to the {\it anchor}. The choice of the anchor falls on the advertiser-defined bid term, as that is the one phrase we know is related to the ad. In rare cases when anchor context vector is not available, one can be constructed as described in Section 3.3. The algorithm uses context vectors of queries learned from search sessions. It initializes the content ad vector to the context vector of the bid term. Next, it creates a textual ad document from all possible $n$-gram phrases contained in the ad title, description, and URL. Finally, it considers all phrases from the ad document for which a context vector exists and has a high enough cosine similarity to the bid term vector (above threshold $\tau_c$). These relevant phrases are then added to the ad vector to further focus its representation. The resulting ad vector lies in the same embedding space as the search queries. Therefore, relevant broad match query candidates can be found via $K$-nearest neighbor search. We refer to this method as {\it content search2vec}.

\begin{table*}[t]
\scriptsize
\caption{Demonstration of elasting search2vec}
\label{tbl:elastic_test}
\begin{center}
\rowcolors{1}{white}{white!10}
\begin{tabular}{lcc}
\rule{0pt}{2.5ex}{\bf User search query (not present in $V_q$)} & {\bf Matched query-document} & {\bf Query-document features (K-nn queries)} \\
\hline \hline
\rowcolor{lightgray}
& & {\bf metropolitan opera house}, ny {\bf opera} \\ 
\rowcolor{lightgray}
metropolitan opera house that is in new york city & metropolitan opera house & {\bf new york opera}, {\bf new york opera house} \\ 
\rowcolor{lightgray}
& & {\bf metropolitan opera}, nyc {\bf opera house} \\ 
& & {\bf best flight} tickets to {\bf paris}, cheap flights to {\bf paris} \\
best flight deals that travel to paris france & best flight tickets to paris & cheap tickets to {\bf paris}, {\bf best flight} fares to {\bf paris} \\
& & cheap airfare {\bf paris france}, {\bf travel to paris france} \\
\rowcolor{lightgray}
& & {\bf best stock ticker} apps, {\bf stock} tracker {\bf app} in {\bf appstore} \\
\rowcolor{lightgray}
what is the best stock ticker trading app in appstore & best stock ticker apps & {\bf best stock} apps in {\bf appstore}, {\bf stock ticker} apps \\
\rowcolor{lightgray}
& & {\bf best trading} apps, real time stocks on the {\bf app} store \\
\bottomrule
\end{tabular}
\end{center}
\end{table*}

\subsection{Cold start query embeddings}
\label{sec:q2vec}
Quality of the query embedding is contingent on the number of times it was observed in the training data. Therefore, even though our scalable training procedure allows us to train vectors for several hundred millions of queries, we are still limited by the frequency of queries in search sessions. Following the procedure from \cite{mikolov2013distributed} we train context vectors for queries that occur at least $10$ times. Considering that frequency of queries in search logs follows a power low distribution \cite{spink2001searching}, a large number of queries will not meet this frequency requirement. They are referred to as {\it tail} queries, as opposed to {\it head} and {\it torso} queries that occupy the portion of the power law curve for which a context vector can be learned. While individually rare, {\it tail} queries make up a significant portion of the query volume. For this reason, these queries have significant potential for advertising revenue. 

Most major search engines pre-compute broad matches for a large number of {\it head} and {\it torso} queries. This table $T_b$ is then cached for fast access and retrieval. The nonexistence or lack of presence in search sessions makes {\it tail} queries much harder to match against ads \cite{broder09online}, as they lack useful contexts of surrounding web activities to help in matching. Matching them to ads solely based on textual information results in matches of low quality, primarily due to the so-called vocabulary mismatch problem \cite{ribeiro2005impedance}. Finally, short text in queries introduces limitations into producing query embeddings based on their content.

To address this problem we propose to expand the head queries from table $T_b$ with $K$ nearest neighbor queries in the learned embedding space and build an {\it inverted index} to be used for efficient matching against tail queries. In the online phase, when a new user query arrives that has no matched ads we use the inverted index to map the query into related previously-seen queries that have ads associated with them. Finally, the user query inherits ads of the top matched query. To illustrate the matching procedure, in Table \ref{tbl:elastic_test} we show examples of tail queries and top matched queries, obtained via inverted index. Bolded words are the ones matched against the user query. We refer to this method as {\it elastic search2vec}.

Previous attempts at building an inverted index of head queries for dealing with tail queries were based on features extracted directly from words of the head query \cite{broder09online} or from words of queries that co-occurred in the same sessions as the head query \cite{broccolo}, such as typos and reformulations. While words from query are good representations of its syntax, they fail to accurately represent its semantics. On the other hand, words from co-occurring queries may not be the best choice as semantically similar queries often have similar contexts (e.g., same link and ad clicks), but do not necessarily appear in the same session. For this reason, we build the inverted index by considering words of semantically similar queries obtained by K-nn search in the embedding space $V_q$. As our experiments confirm, this leads to more accurate cold-start query embeddings and better quality ad matches.

\section{Experiments}
\label{sec:experiments}
In this section we describe the training data set and give empirical evaluation of the proposed search2vec method. The performance was evaluated in terms of relevance on editorially judged set of (query, ad) pairs, as well as in terms of click-through rates from a live bucket tests on Yahoo Search traffic. We compared several variations of our method discussed in the previous sections, as well as a number of baseline approaches. Finally, we conducted experiments to evaluate the proposed models for cold-start ad and query embeddings that address the limitations of the context models by allowing us to produce matches for new queries and ads.

{\bf Search2vec training}. In order to learn query and ad embeddings we organized our search data into sessions, or uninterrupted sequences of user's web search activities: queries, ad clicks, and search link clicks. Each ad impression was uniquely identified by its creative ID and bid term ID. Search sessions that contained only $1$ user action were discarded. We discarded from the vocabulary all activities
that occurred less than $10$ times in the search sessions. The most frequent actions were downsampled with threshold of~$10^{-5}$ as in~\cite{mikolov2013distributed}. The dimensionality of the embedding space was set to $d=300$, the window size was set to $5$ and the number or random negative samples per vector update was set to $5$.

We learned embeddings for more than $126.2$ million unique queries, $42.9$ million unique ads, and $131.7$ million unique links, using one of the largest search data set reported so far, comprising over $9.1$ billion search sessions collected on Yahoo Search. End-to-end search2vec training via parameter server using $10$ iteration over the data set takes about 30 hours to complete. In comparison, multi-threaded single machine implementation \cite{mikolov2013distributed} on 256GB RAM box with 24 threads and same training parameters takes more than a week to complete and can train only up to $80$ million vectors.

\subsection{Broad match models}

{\bf 1) $\text{word2vec}$} model uses publicly available word vectors\footnote{https://code.google.com/p/word2vec, accessed Jan. 2016} trained on Google News data set. Query vectors were generated by summing vectors of word tokens within a query, and ad vectors were generated by summing vectors of word tokens from title, description, bid term, and display URL. English stopwords were not considered in the summation.

{\bf 2) {\bf TF-IDF}} model generates broad match candidates by computing the cosine similarity between bag-of-words representation of queries and ads (constructed from ad title, description, bid term, and display URL, excluding stopwords).

{\bf 3) {\bf $\text{QFG}$}} model was trained using a click-flow graph constructed from search sessions data set $\mathcal{S}$. Broad matches were produced by query rewriting of the ad bid term.

{\bf 4) {\bf search2vec}} model was trained using search sessions data set $\mathcal{S}$ composing of search queries, ads and links. Broad match candidates for a query were generated by calculating cosine similarity between the query vector and all ad vectors.

{\bf 5) $\text{search2vec}_{dwell,in}$} is a variation of search2vec method with dwell time weights, and implicit negatives added to the training data. We combined dwell time and implicit negative enhancements under one framework as both of them aim at reducing the effect of bad clicks. 

{\bf 6) {\bf content search2vec}} is tailored for the cold-start case (i.e. ads that do not appear in search sessions). Broad match candidates are found by calculating cosine similarity between the context query vector the content ad vectors.

{\bf 7) {\bf elastic search2vec}} aims at finding ads for tail queries by matching against an inverted index of context queries that already have ad matches.

Our evaluation did not include topic models such as LDA \cite{blei2003latent} or PLSA \cite{hofmann1999probabilistic}, as a number of researchers \cite{hong2010empirical} found that they obtain suboptimal performance on short texts.

\begin{figure}[t]
\centering
\includegraphics[width=0.38\textwidth]{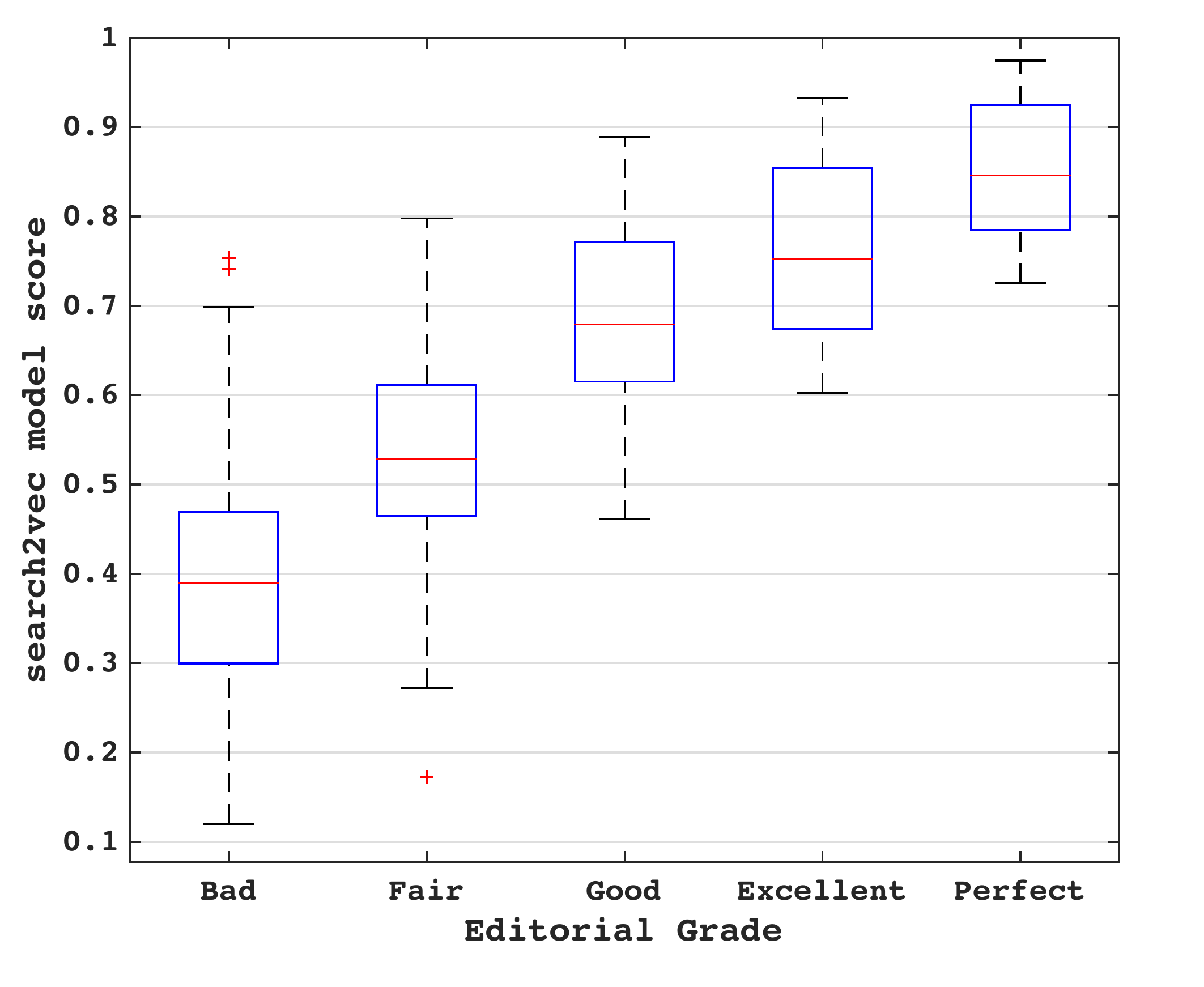}
\caption{Box plot of editorial grades vs. search2vec scores}
\label{fig:antares_edit}
\end{figure}

\subsection{Offline relevance results}
In the first set of experiments we used editorial judgments of query-ad pairs to compare broad match methods in terms of relevance. For this purpose we used an in-house dataset consisting of a query-ad pairs that were graded editorially. The editors were instructed to grade more than $24{,}000$ query-ad pairs as either {\it Bad}, {\it Fair}, {\it Good}, {\it Excellent}, or {\it Perfect} match. For each ad, the editors had access to bid term, ad title, description, and display URL to help them reach their judgment. Note that the candidate query-ad pairs were not generated by any of the tested methods. For each query there were up to 9 judged ads, allowing us to evaluate ranking of ads in addition to relevance.

Using each of the compared methods we generated query features and ad features for each judged pair. Then, to calculate score for a particular pair, we calculated cosine similarity between the query vector and the ad vector. In case of QFG query rewriting model, the query-ad relevance score was calculated between the query and the ad bid term. In case of elastic search2vec model, query vector was obtained by inverse index search that excluded the searched query and calculating the cosine similarity of the top retrieved context query vector and the ad context vector. 

There are several questions one could ask regarding algorithm relevance. First, how well can the algorithm score distinguish between good and bad query-ad pairs. Second, how well can the ads be ranked based on the algorithm score, and how much does this ranking deviate from ranking based on editorial grades. For this purpose we concentrated on the following two metrics to measure relevance:

\begin{figure}[t]
\centering
\includegraphics[width=0.38\textwidth]{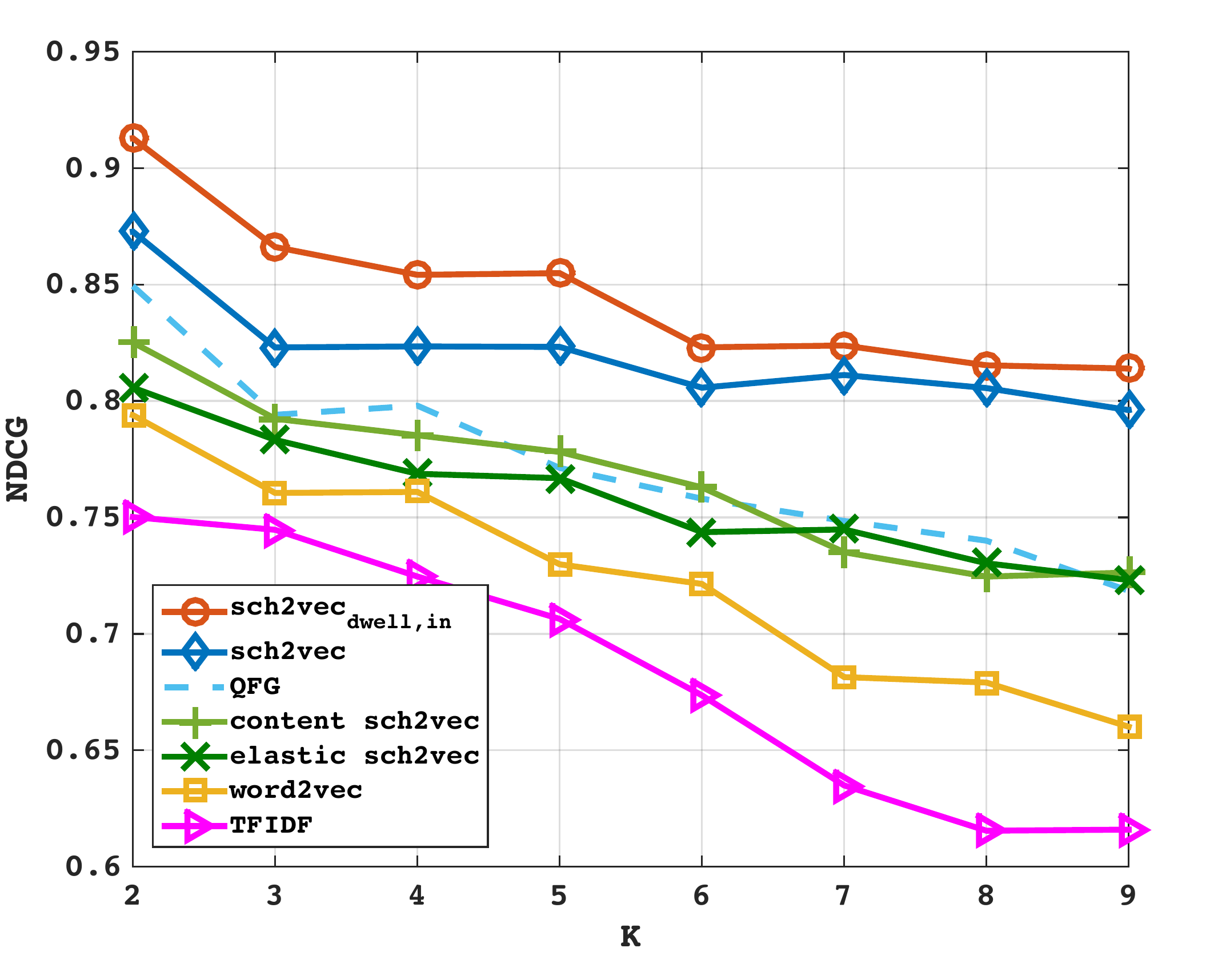}
\caption{NDCG@K for different models}
\label{fig:antares_ndcg}
\end{figure}

\begin{table}[t]
\footnotesize
\caption{Comparison of different methods on editorial data}
\label{tbl:rel_perf}
\begin{center}
\begin{tabular}{lcc}
\rule{0pt}{2.5ex}{\bf Method} & {\bf oAUC} & {\bf Macro NDCG} \\
\hline \hline
\rowcolor{lightgray}
$\text{word2vec}$ & $ 0.6573$ & $0.7308$ \\
$\text{TF-IDF}$ & 0.6407 & $0.6983$ \\ 
\rowcolor{lightgray}
$\text{QFG}$ & 0.6928 & $0.7848$ \\ 
$\text{search2vec}$ & 0.7254 & $0.8303$ \\ 
\rowcolor{lightgray}
{\bf $\text{search2vec}_{dwell,in}$} & ${\bf 0.7392}$ & ${\bf 0.8569}$ \\
$\text{content search2vec}$ & $0.6881$ & $0.7758$ \\ 
\rowcolor{lightgray}
$\text{elastic search2vec}$ & $0.6805$ & $0.7612$ \\ 
\bottomrule
\end{tabular}
\end{center}
\end{table}

{\bf 1) oAUC} Ordinal AUC measures how well the cosine similarity scores computed by the trained model can discriminate between editorial grades, by computing the average AUC (area under ROC curve) across four classifiers: perfect-and-above vs. below-perfect, excellent-and-above vs. below excellent, good-and-above vs. below good, and lastly fair-and-above vs. below fair.

{\bf 2) Macro NDCG} measures how well the ranked scores align with the ranked editorial grades. Given numerical grades (1 for Bad, 2 for Fair, 3 for Good, 4 for Excellent and 5 for Perfect) we used $(2^{grade} - 1)$ as NDCG labels and position discounting of $\log$(position in ranked list). Finally, we average the metric across all queries. Unlike oAUC, this performance measure incurs a greater penalty for mistakes found at the top of the list.

In Table~\ref{tbl:rel_perf} we report the averaged oAUC and Macro NDCG of editorial query-ad pairs for different methods. In addition, to get a sense of how different models compare at a particular rank we show the NDCG metric at values of rank ranging from $2$ to $9$ in Figure~\ref{fig:antares_ndcg}.

Several conclusions can be drawn. First, we can see that the models that did not utilize search sessions (i.e., word2vec and TF-IDF) perform worse than the models that use historical search logs (i.e., QFG and search2vec). Second, search2vec embedding method performs better than QFG graph method. This can be explained by the fact that QFG only makes use of the co-occurrence of user actions in search sessions, while search2vec also accounts for similarity of their contexts. Third, introduction of dwell time and implicit negatives improves relevance over baseline search2vec model.

Interestingly, the increase in oAUC of search2vec$_{dwell, in}$ over search2vec was not as large as the improvement in Macro NDCG. This higher improvement in ranking measure can be explained by the fact that search2vec$_{dwell, in}$ leverages ad views in addition to ad clicks. Therefore, if two ads have the same number of clicks but one of them was viewed an order of magnitude more frequently, search2vec$_{dwell, in}$ will rank it lower than the ad which was shown less times, thus obtaining a better result. Finally, relevance based on cold-start query and ad embeddings produces reasonable results, indicating that they can be successfully used until context vectors can be learned from sessions.

To better characterize the search2vec method, in Figure~\ref{fig:antares_edit} we show a box plot of cosine similarities between editorially judged queries and ads for $\text{search2vec}_{dwell,in}$ model, grouped by different editorial grades. We can see that the cosine similarity scores correlate well with the editorial judgments, indicating a strong potential of the learned embeddings for the task of retrieval of high-quality query-ad pairs.

\subsection{Online test results}
Following successful offline evaluation of the proposed broad match methods, in the second set of experiments we conducted tests on live Yahoo Search traffic. Five buckets with five different broad match techniques were set up (QFG, search2vec, search2vec$_{dwell, in}$, content search2vec and elastic search2vec), all on $5\%$ of overall search traffic and compared against control bucket, which employed a broad match model currently used in production. Note that, although the buckets used different broad match techniques, all of them still included the same exact match and variant match techniques. The online test was run for 7 days, and was performed on a subset of active sponsored search ads.

Different search2vec methods produced broad matches by finding $K=30$ nearest ads in the embedding space for each search query from our $126.2$M vocabulary, and keeping only ads with cosine similarity above $\tau=0.65$. The threshold was chosen based on editorial results from Figure~\ref{fig:antares_edit}, where it can be observed that threshold of $0.65$ captures most Good, Excellent, and Perfect query-ad pairs while discarding most of Fair and Bad ones. While search2vec and search2vec$_{dwell,in}$ were restricted to only ads for which we previously observed clicks in historical search logs, content search2vec generated embeddings for all active ads, and it is expected to have higher ad coverage.
In the case of QFG method, the retrieval was done via query rewriting. For each query we determined the most relevant bid terms, and retrieved all ad creatives with that bid term, up to $K=30$ ads per query.

The results of the test in terms of CTR and coverage are summarized in Table \ref{tbl:ab_test}. For CTR we measured the number of clicks on particular broad match ads, divided by the total number of shown ads by that broad match model. Then, for each broad match model we report a relative improvement of CTR over QFG method, as well as relative change in coverage, defined as a number of ad impressions. 

Considering the results, we can conclude that search2vec$_{dwell, in}$ is the best choice among the competing method as it achieves the highest CTR while maintaining a large coverage. When dwell time and implicit negatives were incorporated in search2vec training we observed a large improvement in CTR, with slight coverage drop, which can be explained by the fact that some ad candidates were eliminated due to low dwell time of high number of skips. Content search2vec and elastic search2vec produced a large increase in coverage since they are not restricted to only clicked ads or queries that appeared in search sessions. At the same time they had satisfactory CTR, indicating that they are  good alternatives to search2vec$_{dwell, in}$.

One of the highlights of the A/B test with the control bucket was that more than $80\%$ of broad matches produced by search2vec$_{dwell, in}$ were unique to that method (i.e., not found by the production model) while retaining high CTR, which can directly lead to incremental gains in revenue. Our search2vec$_{dwell, in}$ combined with cold start ad embeddings was launched in production and it is currently serving more than $30\%$ of all broad matches. 

\begin{table}[t]
\footnotesize
\caption{Comparison of broad match methods in A/B test}
\label{tbl:ab_test}
\begin{center}
\begin{tabular}{lcc}
\rule{0pt}{2.5ex}{\bf Method} & {\bf CTR} & {\bf Coverage} \\
\hline \hline
\rowcolor{lightgray}
$\text{QFG}$ & - & - \\ 
$\text{elastic search2vec}$ & $-4.81\%$ & $+43.79\%$ \\
\rowcolor{lightgray}
$\text{content search2vec}$ & $-1.24\%$ & $+35.61\%$ \\
$\text{search2vec}$ & $+6.21\%$ & $+26.44\%$ \\
\rowcolor{lightgray}
{\bf $\text{search2vec}_{dwell,in}$} & $ +10.64\%$ & $+19.95\%$ \\
\bottomrule
\end{tabular}
\end{center}
\end{table}

\begin{table}[t]
\footnotesize
\caption{Comparison of ad cold-start methods}
\label{tbl:ad_content}
\begin{center}
\begin{tabular}{lcc}
\rule{0pt}{2.5ex}{\bf Method} & {\bf Average} & {\bf STD} \\
\hline \hline
\rowcolor{lightgray}
$\text{bid term vector}$ & 0.731 & 0.128 \\ 
$\text{words}$ & $0.574$ & $0.059$ \\
\rowcolor{lightgray}
$\text{anchor words}$ & $0.688$ & $0.077$ \\
$\text{phrases}$ & $0.665$ & $0.067$ \\ 
\rowcolor{lightgray}
$\text{CRF phrases}$ & $0.604$ & $0.075$ \\ 
{\bf $\text{anchor phrases}$} & {\bf 0.792} & {\bf 0.077 } \\
\rowcolor{lightgray}
\rowcolor{lightgray}

\bottomrule
\end{tabular}
\end{center}
\end{table}

\begin{table}[t]
\footnotesize
\caption{Comparison of query cold-start methods}
\label{tbl:query_content}
\begin{center}
\rowcolors{1}{white}{gray!10}
\begin{tabular}{lcc}
\rule{0pt}{2.5ex}{\bf Method} & {\bf Average} & {\bf STD} \\
\hline \hline
\rowcolor{lightgray}
$\text{words} $ & $0.452$ & $0.101$ \\
$\text{phrases}$ & $0.574$ &$0.120$ \\
\rowcolor{lightgray}
$\text{CRF phrases}$ & $0.514$ &$0.119$ \\ 
$\text{elastic co-occurred queries, K=10}$ & $0.621$ & $0.084$ \\
\rowcolor{lightgray}
$\text{elastic search2vec, K=$5$}$ & $0.685$ & $0.085$ \\
{\bf elastic search2vec, K=10} & {\bf 0.717} & {\bf 0.091} \\
\rowcolor{lightgray}
$\text{elastic search2vec, K=$100$}$ & $0.693$ & $0.089$ \\
\bottomrule
\end{tabular}
\end{center}
\end{table}

\subsection{Cold start experiments}
To evaluate different algorithms for cold start ad embeddings we conducted the following experiment. We used search2vec$_{dwell, in}$ context vectors of queries to construct ad content vectors based on available textual information: title, description, display URL and bid term. To achieve this task, we tested several methods that use different strategies to extract the most relevant keywords from ad text and calculate the embedding as a linear combination of their vectors \eqref{content_vec}. To this end, we evaluated the following techniques: 1) {\it bid term vector}, which uses the vector of a bid term query as the ad vector 2) {\it words}, similar to \cite{le2014distributed} the sum \eqref{content_vec} uses vectors of all words in the document 3) {\it phrases}, which uses vectors of all possible $n$-grams from the ad texr, $n=1,..,10$ 4) {\it CRF phrases}, which uses vectors of phrases obtained by CRF segmentation \cite{lafferty2001conditional} 5) {\it anchor phrases}, which is explained in Algorithm 1 and 6) {\it anchor words}, which is the Algorithm 1 that uses words instead of phrases. In all above-mentioned algorithms, except {\it anchor} algorithms, a hand curated collection of stopwords (e.g., ``free shipping" and ``best prices") were dropped to improve performance. The performance was measured in terms of cosine similarity between the ad context vector and its corresponding content vector. High similarity indicates that the cold-start method can produce a vector that is close enough to the one that would be learned from sessions, and could therefore be matched to relevant queries in the embedding space. In Table~\ref{tbl:ad_content} we show results on a subset of $2$M ads. We can observe that {\it anchor phrases}, which automatically discards uninformative phrases, outperformed the competing baselines by a large margin.

We repeat a similar procedure for queries. Specifically, we test how well the vectors of last $50$M queries in our vocabulary could be generated using vectors of $40$M head queries. It should be noted that generating query content vectors is much more challenging as we have less content (i.e., text) to work with and no concept of {\it anchor} point that could help in extraction. To address this issue we proposed {\it elastic search2vec} described in Section 3.3. We evaluate 3 different sizes of query expansions when creating the inverted index: K$=5$, 10 and 100. In addition to previously mentioned baselines, we also test an inverted index approach similar to \cite{broccolo}, where index is generated based on words from K$=10$ top co-occurring queries. In Table~\ref{tbl:query_content} we summarize the results, where we can observe that the proposed approach with K$=10$ achieved the best performance. 

\subsection{Open-source query embeddings}
As part of this research, we open-sourced $8$M query vectors trained using search2vec\footnote{\url{http://webscope.sandbox.yahoo.com/catalog.php?datatype=l&did=73}}. The vectors may serve as a testbed for query rewriting task as well as to word and sentence similarity task, which are common problems in NLP research. We would like for researchers to be able to produce query rewrites based on these vectors and test them against other state-of-the-art techniques. In addition, we provide an editorially judged set of $4016$ query rewrites, on which we compared search2vec performance against word2vec. The results are summarized in Table~\ref{tbl:open_source}. 

\begin{table}[t]
\footnotesize
\caption{Comparison of query rewriting methods}
\label{tbl:open_source}
\begin{center}
\begin{tabular}{lcc}
\rule{0pt}{2.5ex}{\bf Method} & {\bf oAUC} & {\bf Macro NDCG@5} \\
\hline \hline
\rowcolor{lightgray}
$\text{word2vec}$ & 0.817 & $0.929$\\ 
{\bf search2vec} & {\bf 0.880} & ${\bf 0.959}$ \\
\bottomrule
\end{tabular}
\end{center}
\end{table}

\section{Conclusions and future work}
\label{sec:conclusion}
We proposed a novel broad match method based on neural language models. The method learns low-dimensional representations of search queries and ads based on contextual co-occurrence in user search sessions. To better leverage available web search contexts, we incorporate link clicks, dwell time and implicit negative signals into the training procedure. We evaluated the proposed method using both editorial data and online test on live search traffic. When compared to the baseline approaches, we showed that proposed search2vec model generated the most relevant broad matches between queries and ads, and had the highest CTR. Moreover, we found that in the case of new queries and ads the proposed cold-start embeddings are a good substitute for the learned ones. The results clearly indicate significant advantages of search2vec model over the existing broad match algorithms, and suggest high monetization potential in sponsored search. Finally, to address the scalability issues with regards to data and vocabulary size, we propose a novel distributed algorithm that can achieve 10x training speedup and train 5x larger vocabularies, compared to a single machine algorithm. In our future work we plan to make use of additional search contexts to further improve search2vec ad ranking, including user's age, gender and geo location.

\balance
\bibliographystyle{abbrv}
\bibliography{bibliography_short}

\end{document}